\documentclass[aps,prd,showpacs,nofootinbib,superscriptaddress,preprint,tightenlines]{revtex4}

\usepackage{graphicx}
\usepackage[centertags]{amsmath}
\usepackage{amsfonts}
\usepackage{amssymb}
\usepackage{amsthm}
\usepackage{newlfont}
\usepackage{subfigure}

\def\bea{\begin{eqnarray}}
\def\eea{\end{eqnarray}}
\def\be{\begin{equation}}
\def\ee{\end{equation}}

\usepackage{graphicx}

\def\be{\nopagebreak[3]\begin{equation}}
\def\ee{\end{equation}}
\def\ba{\nopagebreak[3]\begin{eqnarray}}
\def\ea{\end{eqnarray}}

\def\R{\mathbb{R}}

\newcommand{\teta}{\rlap{\lower2ex\hbox{$\,\tilde{}$}}\eta{}}

\def\R{\mathbb{R}}

\newcommand{\f}{\frac}

\usepackage{enumerate}

\usepackage{colordvi}

\usepackage[centertags]{amsmath}
\usepackage{amsfonts}
\usepackage{amssymb}
\usepackage{amsthm}
\usepackage{dsfont}
\usepackage{newlfont}
\usepackage{hyperref}
\def\be{\begin{equation}}
\def\ee{\end{equation}}
\def\dif{\textrm{d}}

\def\f{\frac}

\begin{document}
\preprint{\vbox{\baselineskip=12pt \rightline{IGC-15/02-02}
}}

\title{Loop quantum cosmology of Bianchi IX: \\
Inclusion of inverse triad corrections}
\author{Alejandro Corichi}\email{corichi@matmor.unam.mx}
\affiliation{Centro de Ciencias Matem\'aticas, Universidad Nacional Aut\'onoma de
M\'exico, UNAM-Campus Morelia, A. Postal 61-3, Morelia, Michoac\'an 58090,
Mexico}
\affiliation{Center for Fundamental Theory, Institute for Gravitation and the Cosmos,
Pennsylvania State University, University Park
PA 16802, USA}
\author{Asieh Karami}
\email{karami@ipm.ir}
\affiliation{Centro de Ciencias Matem\'aticas, Universidad Nacional Aut\'onoma de
M\'exico, UNAM-Campus Morelia, A. Postal 61-3, Morelia, Michoac\'an 58090,
Mexico}
\affiliation{School of Astronomy, Institute for Research in Fundamental Sciences (IPM), P. O. Box 19395-5531, Tehran, Iran}

\begin{abstract}
We consider the loop quantization of the (diagonal) Bianchi type IX cosmological model. We explore different quantization prescriptions that extend the work of Wilson-Ewing and Singh. In particular, we study two different ways of implementing the so-called inverse triad corrections. We construct the corresponding Hamiltonian constraint operators and show that the singularity is formally resolved.
We find the effective equations associated with the different quantization prescriptions, and study the relation with the
isotropic $k$=1 model that, classically, is contained within the Bianchi IX model. We use
geometrically defined scalar observables to explore the physical implications of each of these theories. This is the first part in a series of papers analyzing different aspects of the Bianchi IX model, with inverse corrections, within loop quantum cosmology.
\end{abstract}

\maketitle

\section{Introduction}
\label{sec:1}

Loop quantum cosmology (LQC) represents an attempt to understand  physics of the early, Planck scale universe by considering seriously the quantum features of the gravitational field. It is based on canonical quantization methods of symmetry reduced general relativity. The main difference with previous attempts being that the quantization strategy follows the one behind loop quantum gravity. The end result is that new effects that arise from the quantum nature of geometry at the Planck scale become important in a certain regime and prevent the classical singularity, replacing it with a `bounce'. For details see \cite{lqc} and for a summary see Ashtekar's contribution to this volume.

The best understood models within LQC are homogeneous and isotropic models, and in particular the $k$=0 FLRW model, 
where a complete quantization has been constructed (see for instance \cite{aps2,slqc,CM,GS}). A physical Hilbert space 
was constructed and the numerical evolution of rational physical observables exhibited a bounce replacing the big bang 
\cite{aps2}. Furthermore, the model was exactly solved and matter density was shown to be absolutely bounded by a 
`critical density' $\rho_c$ of the order of Planck density \cite{slqc}. Furthermore, the dynamics of
semiclassical states can be described by a simple ``effective Hamiltonian" generating effective equations that capture 
the main (loop) quantum gravity corrections to the classical equations of motion \cite{aps2}. It turns out that all 
solutions to the effective equations bounce when the density is precisely $\rho_c$. Analytical and numerical studies 
have shown that semiclassical state follow the effective dynamics and bounce with a density arbitrarily close to $
\rho_c$ \cite{CM,GS}. Furthermore, the quantization prescription that allowed to obtain all these results 
\cite{aps2,slqc} was shown to be unique when consistency and physical criteria are imposed \cite{cs:unique}.
The closed model has also received much attention \cite{closed,CK,CK-2,sv}. Numerical simulation have again shown that 
the big crunch and big bang singularities are replaced by a cyclic universe \cite{closed}, well described by an effective theory. Singularity resolution, just as in the flat case \cite{cosmos}, was shown to be generic \cite{sv}.

The next step within homogenous cosmological models are anisotropic, ``Bianchi" models. some of them are natural extensions of the FLRW isotropic cases. For instance, the Bianchi I model is spatially flat and reduces to the $k$=0 FLRW in its
isotropic sector. The Bianchi IX model, on the other hand, reduces to the $k$=1 FLRW model. This Bianchi IX model is also important for a different reason. The so called BKL conjecture states that for generic inhomogenous models, in the dynamics close to the singularity, time derivatives dominate over space derivatives in such a way that the dynamics of nearby points decouple, and each one behaves as a Bianchi IX model \cite{BKL1,BKL2,BKL-AHS}. The dynamics of
Bianchi IX is interesting by itself, with a `mixmaster' regime that can be described as a series of Bianchi I epochs connected by Bianchi II transitions \cite{wain_ellis}.

A natural question is whether loop quantum cosmology can say something about the singularity resolution and possible modifications to the BKL dynamics near the Planck scale. Of course, the study of anisotropic models is not new in LQC. The Bianchi I model was the first one to be studied \cite{bianchiold,bianchiI,singh-BI}, and the Bianchi II model followed \cite{bianchiII,CM-bianchi2}. The Bianchi IX model, in the simplest case with $N$=$V$ and no inverse corrections,
was first introduced in \cite{Ed-BIX}. Some of this corrections were introduced and studied in \cite{singh-Ed,gupt:singh}.
The issue of ambiguities in the quantum theory is not strange to loop quantization in cosmology. 
While for the simplest $k$=0 FLRW model the quantization is assentially unique 
\cite{cs:unique}, it was first realized in \cite{bianchiII} that a spatially curved anisotropic model 
force to change the quantization strategy to define curvature. Instead of using closed loops, as 
in the isotropic models, one can only define connections though open paths. This ambiguity was 
explored in the
closed $k$=1 FLRW model, where it was shown that the new quantization yields not one but two different bounces \cite{CK,CK-2}. For isotropic models there are also different ways
of introducing the lapse function and the so called inverse corrections. The purpose of this manuscript and those that follow, is to study the ambiguities in the Bianchi IX model by exploring different quantizations.  This paper is the first in a series. Here we introduce the new quantization where due care is taken for the inverse corrections. In the second paper in the series \cite{CM-BIX}, we explore numerically the effective equations here found, for a massless scalar field. In the third paper of the series, we shall explore some qualitative implications of the quantization here presented, including the vacuum case \cite{CKM-2} (some early result have already been presented in \cite{CKM}).

The structure of the manuscript is as follows. In Sec.~\ref{sec:2} we introduce the model and the preliminaries necessary for the rest of the manuscript. In Sec.~\ref{sec:3} we introduce the new quantizations. Sec.!\ref{sec:4} is devoted to the study of the effective equations that arise from the quantum theories defined before. We end in Sec.~\ref{sec:5} with a discussion.

\section{Classical Theory}
\label{sec:2}

Bianchi models are spatially homogeneous models such that the symmetry group $\mathcal S$ acts simply and transitively on the space manifold $\Sigma\cong\mathcal S$. the symmetry group for Bianchi IX model are the three spatial rotations on a 3-sphere. To define fiducial frames and co frames, we identify this group with SU(2) which carries a Cartan connection
\[
{}^o\omega=g^{-1}\dif g={}^o\omega^i\tau_i
\]
This connection satisfies Maurer-Cartan structure equation
$$
\dif {}^o\omega^i+\frac{1}{2}{}^o\epsilon^i_{\ jk}{}^o\omega^j\wedge{}^o\omega^k=0
$$
Where ${}^o\epsilon_{ijk}$ is the completely antisymmetric tensor and defined such that ${}^o\epsilon_{123}=1$.
We denote dual vectors ${}^oe_i$ corresponding to ${}^o\omega^i$ such that ${}^oe_i^a{}^o\omega^j_a=\delta_i^j$ and ${}^oe_i^a{}^o\omega^i_b=\delta_b^a$. These vectors satisfy the Lie bracket
$$
[{}^oe_i,{}^oe_j]={}^o\epsilon_{ij}^{\ \ k}{}^oe_k.
$$
Therefore the fiducial metric on $\Sigma$ is
$${}^o\!q_{ab}:={}^o\!\omega^i_a\,{}^o\!\omega^j_b\,k_{ij},$$
with $k_{ij}$ the Killing-Cartan metric on su(2). This fiducial metric is the metric of a 3-sphere with radius $a_o=2$. The volume of this 3-sphere is $V_o=2\pi^2\,a_0^3$. It is useful to define $\ell_o=V_o^{1/3}$ and $\vartheta=\ell_o/a_o$.

In general relativity in Ashtekar-Barbero variables, the gravitational phase space consists of pairs $(A_a^i,E_i^a)$ on $\Sigma$ where $A_a^i$ is a SU(2) connection and $E_i^a$ is a densitized triad of weight 1. Since the Bianchi IX model is homogeneous and, if we restrict ourselves to diagonal metrics, one can fix the gauge in such a way that
$A_a^i$ has 3 independent components, $c^i$, and $E_i^a$ has 3 independent components, $p_i$,
\begin{equation}
A_a^i=\frac{c^i}{\ell_o}{}^o\omega_a^i\ \ \ \textrm{and}\ \ \ E_i^a=\frac{p_i}{\ell_o^2}\sqrt{{}^oq}{}\ ^oe_i^a
\end{equation}
where $p_i$ in terms of the scale factors $a_i$ are $|p_i|=\ell_o^2a_ja_k$ ($i\neq j\neq k$). $c_i$ are dimension-less and $p_i$ have dimensions of length-squared.
Using $(c^i,p_i)$ the Poisson brackets can be expressed as
$$\{c^i,p_j\}=8\pi G\gamma\delta_j^i$$
where $\gamma$ is the Barbero-Immirizi parameter.
The physical frames and co-frames are
\be
\omega^i=a^i{}^o\!\omega^i\ \ \ \textrm{and}\ \ \ a_ie_i={}^o\!e_i.
\ee
The physical metric in diagonal manner can be written as
\be
q_{ab}=a_i^2{}^o\!\omega_a^i{}^o\!\omega_b^i.
\ee
and thus the physical volume of $\Sigma$ is $V=2\pi^2a_1a_2a_3$ which is equal to $\sqrt{|p_1p_2p_3|}$.

Since the fiducial frames and co-frames are fixed and because of the parametrization of connections and triads, the only relevant constraint is the Hamiltonian constraint that has the form,
\be
\mathcal{C}_H=N\bigg(\mathcal{H}_{\textrm{grav}}+\mathcal{H}_{\textrm{matter}}\bigg),
\ee
where $N$ is the lapse function, $\mathcal{H}_{\textrm{matter}}$ is $\rho V$ ($\rho$ is 
the matter density) and
\be
\label{hgc}
{\cal H}_{\textrm{grav}}=\int_\mathcal V \left[-\frac{\epsilon^{ij}_{\ k}E_i^aE_j^b}{16\pi G\gamma^2\sqrt{|q|}}\left(
F_{ab}^k-(1+\gamma^2)\Omega_{ab}^k\right)\right] ,
\ee
where $e=\sqrt{|\textrm{det}E|}$, $F_{ab}^k$  and $\Omega_{ab}^k$ are respectively the curvature of connection $A_a^i$ and the curvature of the spin-connection $\Gamma^i_a$ compatible with the triad.
$F_{ab}^k$ in terms of phase space variables is
\be
F_{ab}^k=\f{2}{\ell_o^2}(\varepsilon c_ic_j-2\vartheta c_k){}^o\!\epsilon^{ijk}{}^o\!\omega_{a}^i{}^o\!\omega_{b}^j
\ee
where $\varepsilon$ shows the orientation of physical frames (that is, $\varepsilon=$ 1 when $p_1p_2p_3\geq0$ and $\varepsilon=$-1 when $p_1p_2p_3<0$).

For calculating the spin connection curvature it is convenient to first compute $\Gamma^i$.
\be
\Gamma^i_a=\f{\varepsilon}{a_o}\bigg(\f{a_j}{a_k}+\f{a_k}{a_j}-\f{a_i^2}{a_ja_k}\bigg){}^o\!\omega_a^i
=\f{\varepsilon}{a_o}\bigg(\f{p_j}{p_k}+\f{p_k}{p_j}-\f{p_jp_k}{p_i^2}\bigg){}^o\!\omega_a^i\ \ \ \ i\neq j\neq k.
\ee
and then
\be
\Omega_{ab}^{k}=-\f{2\varepsilon}{a_o^2}\bigg(3\f{p_ip_j}{p_k^2}+2\f{p_k^2}{p_ip_j}-2\f{p_i}{p_j}-2\f{p_j}{p_i}-\f{p_k^2p_i}{p_j^3}-\f{p_k^2p_j}{p_j^3}\bigg)
{}^o\!\omega_a^k\ \ \ \ i\neq j\neq k.
\ee
So the classical Hamiltonian constraint is given by
\be
\begin{split}
\mathcal C_H=&-\frac{N}{8\pi G\gamma^2\lambda^2}\bigg(\textrm{sgn}(p_1p_2)\sqrt{\bigg|\f{p_1p_2}{p_3}\bigg|}c_1c_2
+\textrm{sgn}(p_2p_3)\sqrt{\bigg|\f{p_2p_3}{p_1}\bigg|}c_2c_3+\textrm{sgn}(p_1p_3)\sqrt{\bigg|\f{p_1p_3}{p_2}\bigg|}c_1c_3\\
&-\vartheta\bigg[\textrm{sgn}(p_1\sqrt{\bigg|\f{p_2p_3}{p_1}\bigg|}c_1+\textrm{sgn}(p_2)\sqrt{\bigg|\f{p_1p_3}{p_2}\bigg|}c_2
+\textrm{sgn}(p_3)\sqrt{\bigg|\f{p_1p_2}{p_3}\bigg|}c_3\bigg]\\
&+\vartheta^2(1+\gamma^2)\bigg[2|p_1|\sqrt{\bigg|\f{p_1}{p_2p_3}\bigg|}+2|p_2|\sqrt{\bigg|\f{p_1}{p_1p_3}\bigg|}+2|p_3|\sqrt{\bigg|\f{p_3}{p_1p_2}\bigg|}\\
&-\f{|p_1p_2|^{3/2}}{|p_3|^{5/2}}-\f{|p_2p_3|^{3/2}}{|p_1|^{5/2}}-\f{|p_1p_3|^{3/2}}{|p_2|^{5/2}}\bigg]\bigg)\\
&+\rho\sqrt{|p_1p_2p_3|}
\end{split}
\ee

In the rest of the manuscript, we choose lapse function $N$ to be equal to 1.\footnote{This choice will allow us to include more corrections to the effective Hamiltonian in Sec. \ref{sec03}.}

\section{Quantum Theory}
\label{sec:3}

To construct the quantum kinematics, we have to select a set of elementary observables such that their associated operators are unambiguous. In loop quantum gravity they are the holonomies $h_e$ defined by the connection $A_a^i$ along edges $e$ and the fluxes of the densitized triad $E_i^a$ across surfaces \cite{acz}.
For our model we choose $p_i$ and $e^{i\mu c_i}$ (because a holonomy along the edge $e_i$ parallel to $i$-th vector basis with length $\mu$ is made by the combination of these operators).
We generate the gravitational part of the kinematical Hilbert space by considering  countable linear combinations of orthonormal basis $\{|l_1,l_2,l_3\rangle:\ l_1,l_2,l_3\in\R\}$, where in this basis the operators $\hat p_i$'s are diagonalized and satisfy
\be
\langle l_1,l_2,l_3|l^\prime_1,l^\prime_2,l^\prime_3\rangle=\delta_{l_1,l^\prime_1}\delta_{l_2,l^\prime_2}\delta_{l_3,l^\prime_3}.
\ee.
The elements of this space are then square summable functions.

The action of the elementary operators on this basis are
$$\hat p_i|l_1,l_2,l_3\rangle=(2V_c)^{2/3}\textrm{sgn}(l_i)l_i^2|l_1,l_2,l_3\rangle$$
and
$$e^{i\mu c_i}|l_1,l_2,l_3\rangle=|l_i-\f{\textrm{sgn}(\mu)\sqrt{\mu|}}{1v_c^{1/3}},l_j,l_k\rangle\quad,\quad i\neq j\neq k$$
where $V_c=2\pi G\hbar\gamma\lambda$.


To have the corresponding constraint operator, one needs to express it in terms of the chosen phase space functions $e^{i\mu c_i}$ and $p_i$.
The first term, $\epsilon^{ij}_{\ k}E_i^aE_j^b/\sqrt{|q|}$, as in loop quantum gravity, can be treated by using Thiemann's strategy \cite{TT}.
\begin{equation}
\epsilon_{ijk}\frac{E^{ai}E^{bj}}{\sqrt{|q|}}=\sum_i\frac{1}{2\pi\gamma G\mu}\ ^o\epsilon^{abc}\ ^o\omega_c^i\textrm{Tr}(h_i^{(\mu)}\{h_i^{(\mu)-1},V\}\tau_k)
\label{ths}
\end{equation}
where $h_i^{(\mu)}$ is the holonomy along the edge parallel to $i$-th vector basis with length $\mu$ and $V$ is the volume, which is equal to $\sqrt{|p_1p_2p_3|}$. Note that $\mu$ is arbitrary.
Now, to define an operator related to the first term of Eq.(\ref{hgc}), we can use the right hand side of
Eq.(\ref{ths}) and replace Poisson brackets with commutators.
To find an operator related to the curvature $F_{ab}^k$, for isotropic models and Bianchi I, one can consider a square
$\square_{ij}$ in the $i-j$ plane which is spanned by two of the fiducial triads (for the closed isotropic model since
triads do not commute, to define this plane we use a triad and a right invariant vector ${}^o\xi_i^a$),
with each of its sides having length $\mu_i^\prime$. Therefore, $F_{ab}^k$ is given by
\begin{equation}
F_{ab}^k=2\lim_{Area_\square\rightarrow 0}\epsilon_{ij}^{\ \ k}\textrm{Tr}\bigg(\frac{h_{\square_{ij}}^{\mu^\prime}-\mathbb I}{\mu^\prime_i\mu^\prime_j}\tau^k\bigg){}^o\omega_a^i{}^o\omega_b^j \, .
\label{fs}
\end{equation}
Since in loop quantum gravity, the area operator does not have a zero eigenvalue, one can take the limit of Eq.(\ref{fs})
to the point where the area is
equal to the smallest eigenvalue of the area operator, $\lambda^2 =4 \sqrt{3} \pi \gamma l_{p}^2$,
instead of zero. Then, $\mu_i^\prime a_i=\lambda$. We take  $\mu_i^\prime=\bar\mu_i\ell_o$ where $\bar\mu_i$ is a dimensionless parameter and, by previous considerations,  is equal to $\bar\mu_i=\lambda\sqrt{|p_i|}/\sqrt{|p_jp_k|}$ ($i\neq j\neq k$).

For Bianchi IX, we cannot use this method because the resulting operator is not almost periodic, therefore
we express the connection $A_a^i$ in terms of holonomies and then use the standard definition of curvature $F_{ab}^k$ \cite{bianchiII,Ed-BIX}.
$$
A_a^i=\lim_{\ell_i\rightarrow 0}\frac{1}{2\ell_i}(h^{(\ell_i)}-h^{(\ell_i)-1})
$$
To be consistent with other models, we choose
$$\ell_i=2\mu^\prime_i$$
Thus the operators corresponding to the connection are given by \cite{bianchiII,Ed-BIX}
\begin{equation}
\hat c_i=\widehat{\frac{\sin\bar\mu_ic_i}{\bar\mu_i}}\, .
\end{equation}

Also, one can see that the terms related to the curvatures,
$F_{ab}^k$ and $\Omega_{ab}^k$, contain some negative powers of $p_i$ which are not well defined operators. To solve this problem we use the same idea in Thiemann's strategy.
\begin{equation}
|p_i|^{(\ell-1)/2}=-\frac{\sqrt{|p_i|}\ell_o}{2\pi G\gamma\tilde\mu_i\ell}\textrm{Tr}(\tau_i h_i^{(\tilde\mu_i)}\{h_i^{(\tilde\mu_i)-1},|p_i|^{\ell/2}\}) \, ,
\label{np}
\end{equation}
where $\tilde\mu_i$ is the length of a curve and $\ell \in (0,1)$.
Therefore, for these three different operators we have three different curve lengths ($\mu,\mu^\prime,\tilde\mu$) where $\mu$ and
$\tilde\mu$ can be some arbitrary functions of $p_i$. For simplicity
we shall choose all of them to be equal to $\mu^\prime$. On the other hand we have another free parameter in the definition of
negative powers of $p_i$ which is $\ell$. Since the largest negative power of $p_i$
which appears in the constraint is $-1/4$, we will take $\ell=1/2$ and obtain it directly from Eq.(\ref{np}),
and after that we express the other negative powers in terms of them.

By the above choices, the operators related to the Eqs.(\ref{ths}, \ref{np}) take the form
\bea
\widehat{\epsilon^{ij}_{\ k}\frac{E_i^aE_j^b}{\sqrt{|q|}}}&=&\frac{\hat\epsilon\widehat{\sqrt{\bigg|\f{p_ip_j}{p_k}\bigg|}}}{4\pi G\gamma\lambda}(\widehat{e^{i\bar\mu_k c_k}}\hat V\widehat{e^{-i\bar\mu_k c_k}}-\widehat{e^{-i\bar\mu_k c_k}}\hat V\widehat{e^{i\bar\mu_k c_k}})\ ^o\epsilon^{abc} \ ^o\omega^k_c\nonumber\\
&&=\widehat{\sqrt{\bigg|\f{p_ip_j}{p_k}\bigg|}}\hat A_k\ ^o\epsilon^{abc} \ ^o\omega^k_c\, ,
\eea
and
\be
\widehat{|p_i|^{-1/4}}=\frac{\sqrt{|p_jp_k|}}{2\pi G\gamma\lambda}(\widehat{e^{i\bar\mu_i c_i/2}}\hat {|p|}^{1/2}\widehat{e^{-i\bar\mu_i c_i/2}}-\widehat{e^{-i\bar\mu_i c_i/2}}\hat {|p|}^{1/2}\widehat{e^{i\bar\mu_i c_i/2}}).
\ee
where
\be\label{Adb}
\hat A_k=\frac{\hat\epsilon}{4\pi G\gamma\lambda}(\widehat{e^{i\bar\mu_k c_k/2}}\hat V\widehat{e^{-i\bar\mu_k c_k/2}}-\widehat{e^{-i\bar\mu_k c_k/2}}\hat V\widehat{e^{i\bar\mu_k c_k/2}})
\ee

In the above equations instead of $e^{i\mu c_i}$ with arbitrary real number $\mu$, the operators $e^{i\bar\mu_i c_i/2}$ and their powers appear. Therefore we choose $e^{i\bar\mu_i c_i/2}$ to be the elementary operators along with $p_i$. The action of these operators are given by
$$e^{i\bar\mu_i c_i}|l_1,l_2,l_3\rangle=|l_i-\f{1}{|l_jl_k|},l_j,l_k\rangle\quad,\quad i\neq j\neq k.$$

Also, since the operators $\hat A_k$ have the same action on elements of Hilbert space, in the rest we
denote them by $\hat A$.

Using these results, the constraint operator without factor ordering is
\be
\begin{split}
\hat{\mathcal C_H}=&-\frac{1}{8\pi G\gamma^2\lambda^2}\hat A\widehat{p_1^{-1/2}}\widehat{p_2^{-1/2}}\widehat{p_3^{-1/2}}(\hat{p_1}\hat{p_2}\hat{|p_3|}\sin\bar\mu_1c_1\sin\bar\mu_2c_2\\
&+\hat{p_1}\hat{|p_2|}\hat{p_3}\sin\bar\mu_1c_1\sin\bar\mu_3c_3
+\hat{|p_1|}\hat{p_2}\hat{p_3}\sin\bar\mu_2c_2\sin\bar\mu_3c_3)\\
&+\frac{\vartheta}{4\pi G\gamma^2\lambda}\hat\varepsilon\hat A\bigg(\hat{p_1}\hat{p_2}\widehat{p_3^{-1}}\sin\bar\mu_3c_3+
\hat{p_2}\hat{p_3}\widehat{p_1^{-1}}\sin\bar\mu_1c_1+\hat{p_1}\hat{p_3}\widehat{p_2^{-1}}\sin\bar\mu_2c_2\bigg)\\
&-\frac{\vartheta^2(1+\gamma^2)}{8\pi G\gamma^2}\hat A\bigg(2\widehat{p_1^{3/2}}\widehat{p_2^{-1/2}}\widehat{p_3^{-1/2}}+2\widehat{p_2^{3/2}}\widehat{p_1^{-1/2}}\widehat{p_3^{-1/2}}
+2\widehat{p_3^{3/2}}\widehat{p_1^{-1/2}}\widehat{p_2^{-1/2}}\\
&-\widehat{p_1^{3/2}}\widehat{p_2^{3/2}}\widehat{p_3^{-5/2}}-\widehat{p_1^{3/2}}\widehat{p_3^{3/2}}\widehat{p_2^{-5/2}}
-\widehat{p_2^{3/2}}\widehat{p_3^{3/2}}\widehat{p_1^{-5/2}}\bigg)\\
&+\hat\rho\widehat{p_1^{1/2}}\widehat{p_2^{1/2}}\widehat{p_3^{1/2}}
\end{split}
\ee

After choosing some factor ordering, we can construct the total constraint operator. Note that  different choices of factor
ordering will yield different operators, but the main results will remain almost the same. By solving the constraint equation
$\hat{\mathcal C}_H\cdot\Psi=0$, we can obtain the physical states and the physical Hilbert space $\mathcal H_{\textrm phys}$.
As a final step, one would need to identify the physical observables, that in our case would 
correspond to relational observables as functions of the internal time $\phi$.
Here we choose the factor ordering which is similar to the one used in \cite{bianchiI,bianchiII,Ed-BIX}.

\be
\hat{\mathcal C}_H =\hat{\mathcal C}^{(1)}+\hat{\mathcal C}^{(2)}+\hat{\mathcal C}^{(3)}+\hat{\mathcal H}_{\textrm{matt}}
\ee
where
\be
\hat{\mathcal C}^{(1)}=\frac{1}{32\pi G\gamma^2\lambda^2}\sum_{i=1}^3\sum_{j\neq i}^3\hat{\mathcal C}^{(1)++}_{ij}+\hat{\mathcal C}^{(1)+-}_{ij}+\hat{\mathcal C}^{(1)-+}_{ij}+\hat{\mathcal C}^{(1)--}_{ij}
\ee

\bea
\hat{\mathcal C}^{(1)++}_{ij}&=&\bigg(\frac{\hat V}{\hat{\sqrt V}}\bigg)^{2/3}(\textrm{sgn}(l_i)e^{i\bar\mu_ic_i}+e^{i\bar\mu_ic_i}\textrm{sgn}(l_i))\bigg(\frac{\hat V}{\hat{\sqrt V}}\bigg)^{2/3}\nonumber\\
&&\times\hat A (\textrm{sgn}(l_j)e^{i\bar\mu_jc_j}+e^{i\bar\mu_jc_j}\textrm{sgn}(l_j))\bigg(\frac{\hat V}{\hat{\sqrt V}}\bigg)^{2/3}\nonumber\\
\hat{\mathcal C}^{(1)+-}_{ij}&=&-\bigg(\frac{\hat V}{\hat{\sqrt V}}\bigg)^{2/3}(\textrm{sgn}(l_i)e^{-i\bar\mu_ic_i}+e^{-i\bar\mu_ic_i}\textrm{sgn}(l_i))\bigg(\frac{\hat V}{\hat{\sqrt V}}\bigg)^{2/3}\nonumber\\
&&\times\hat A (\textrm{sgn}(l_j)e^{i\bar\mu_jc_j}+e^{i\bar\mu_jc_j}\textrm{sgn}(l_j))\bigg(\frac{\hat V}{\hat{\sqrt V}}\bigg)^{2/3}\nonumber\\
\hat{\mathcal C}^{(1)-+}_{ij}&=&- \bigg(\frac{\hat V}{\hat{\sqrt V}}\bigg)^{2/3}(\textrm{sgn}(l_i)e^{i\bar\mu_ic_i}+e^{i\bar\mu_ic_i}\textrm{sgn}(l_i))\bigg(\frac{\hat V}{\hat{\sqrt V}}\bigg)^{2/3}\nonumber\\
&&\times\hat A (\textrm{sgn}(l_j)e^{-i\bar\mu_jc_j}+e^{-i\bar\mu_jc_j}\textrm{sgn}(l_j))\bigg(\frac{\hat V}{\hat{\sqrt V}}\bigg)^{2/3}\nonumber\\
\hat{\mathcal C}^{(1)--}_{ij}&=&\bigg(\frac{\hat V}{\hat{\sqrt V}}\bigg)^{2/3}(\textrm{sgn}(l_i)e^{-i\bar\mu_ic_i}+e^{-i\bar\mu_ic_i}\textrm{sgn}(l_i))\bigg(\frac{\hat V}{\hat{\sqrt V}}\bigg)^{2/3}\nonumber\\
&&\times\hat A (\textrm{sgn}(l_j)e^{-i\bar\mu_jc_j}+e^{-i\bar\mu_jc_j}\textrm{sgn}(l_j))\bigg(\frac{\hat V}{\hat{\sqrt V}}\bigg)^{2/3}\nonumber\\
&&
\eea

\bea
\hat{\mathcal C}^{(2)}=&-&\frac{i\vartheta}{16\pi G\gamma^2\lambda}\bigg[\hat p_2\hat p_3\hat{|p_1|}^{-1/2}\bigg(\hat\epsilon\hat A(e^{i\bar\mu_1c_1}-e^{-i\bar\mu_1c_1})+(e^{i\bar\mu_1c_1}-e^{i\bar\mu_1c_1})\hat\epsilon\hat A\bigg)\hat{|p_1|}^{-1/2}
\nonumber\\
&+& \hat p_1\hat p_3\hat{|p_2|}^{-1/2}\bigg(\hat\epsilon\hat A(e^{i\bar\mu_2c_2}-e^{-i\bar\mu_2c_2})+(e^{i\bar\mu_2c_2}-e^{i\bar\mu_2c_2})\hat\epsilon\hat A\bigg)\hat{|p_2|}^{-1/2}
\nonumber\\
&+&\hat p_1\hat p_2\hat{|p_3|}^{-1/2}\bigg(\hat\epsilon\hat A(e^{i\bar\mu_3c_3}-e^{-i\bar\mu_3c_3})+(e^{i\bar\mu_3c_3}-e^{i\bar\mu_3c_3})\hat\epsilon\hat A\bigg)\hat{|p_3|}^{-1/2}\bigg]
\eea

\bea
\hat{\mathcal C}^{(3)}&=&-\frac{\vartheta^2(1+\gamma^2)}{8\pi G\gamma^2}\hat A\bigg(2\widehat{p_1^{3/2}}\widehat{p_2^{-1/2}}\widehat{p_3^{-1/2}}+2\widehat{p_2^{3/2}}\widehat{p_1^{-1/2}}\widehat{p_3^{-1/2}}
+2\widehat{p_3^{3/2}}\widehat{p_1^{-1/2}}\widehat{p_2^{-1/2}}\nonumber\\
&&-\widehat{p_1^{3/2}}\widehat{p_2^{3/2}}\widehat{p_3^{-5/2}}-\widehat{p_1^{3/2}}\widehat{p_3^{3/2}}\widehat{p_2^{-5/2}}
-\widehat{p_2^{3/2}}\widehat{p_3^{3/2}}\widehat{p_1^{-5/2}}\bigg)
\eea
and by choosing massless scalar field (as internal time), the matter part is given by
\be
\hat{\mathcal H}_{\textrm{matt}}=\f{1}{2}\hat{p}_\phi^2\,\widehat{p_1^{-1}}\widehat{p_1^{1/2}}\widehat{p_2^{-1}}\widehat{p_2^{1/2}}\widehat{p_3^{-1}}\widehat{p_3^{1/2}}
\ee

To calculate the action of the constraint operator it is simpler to work with dimensionless variable 
$v$, which is related to the volume, and two variables of three $l_1$, $l_2$ and $l_3$. 
The quantity $v$ is equal to $2l_1l_2l_3$ and $\hat V=V_c|v||l_1,l_2,l_3\rangle$. Because of the 
symmetry in the model, there is no preference to choose one of $l_i$'s to replace with $v$. Here, 
we choose $l_3$. One should note that we cannot use variable $v$ in a case that the state has zero volume. 
However, it is easy to see that the constraint operator annihilates the states with zero volume and also 
the other states cannot reach those states. Therefore, the use of variable $v$ is fully justified.

In those variables, the action of operators $e^{i\bar\mu_ic_i/2}$, $\hat A$ and $\widehat{|p|^{-1/4}}$ are given by
\be
e^{i\bar\mu_1c_1/2}|l_1,l_2,v\rangle=|l_1-\f{1}{|l_1v|},l_2,v-\textrm{sgn}(l_1v)\rangle
\ee
and
\be
\hat A|l_1,l_2,v\rangle=A(v)|l_1,l_2,v\rangle,
\ee
and
\be
\widehat{|p|^{-1/4}}|l_1,l_2,v\rangle=\frac{h(v)}{V_c}\prod_{j\neq i}p_j^{1/4}|l_1,l_2,v\rangle
\ee
where
\bea
\label{Avl}
A(v)&=&(||v|+1|-||v|-1|)=
\left\{\begin{array}{lr}|v| & |v|<1\\1 & |v|\geq 1\end{array}\right.,\\
h(v)&=&\sqrt{V_c}\bigg(\sqrt{||v|+1|}-\sqrt{||v|-1|}\bigg).
\label{hvl}
\eea

In \cite{CK-2} we showed that the operators similar to the constraint operator for closed FLRW model are
essentially self adjoint and since, here, the constraint operator has a similar form as the FLRW one, it is reasonable to expect it to be essentially self adjoint, too; thus we will work on its extended domain.

Also, since the constraint operator is invariant under parity, to see the full action of this operator on a state, one just needs to calculate its action on the positive octant (which means $l_1,l_2,v>0$). The action of the constraint operator on state $\Psi(l_1,l_2,v;\phi)$ is then given by
\bea
-\partial^2_\phi\Psi(l_1,l_2,v;\phi)&=&\f{V_cv^{-5}h^{-12}(v)}{8\pi G\gamma}\Bigg[\f{\chi_{v-4}V_cv^{4/3}}{2\lambda^2}(v-2)^{4/3}(v-4)^{4/3}A(v-2)\nonumber\\
&&\times h^{4/3}(v)h^{4/3}(v-2)h^{4/3}(v-4)\Psi_{-4}(l_1,l_2,v;\phi)\nonumber\\
&&+\f{1}{2\lambda^2}V_cv^{4/3}(v+2)^{4/3}(v+4)^{4/3}A(v+2)\nonumber\\
&&\times h^{4/3}(v)h^{4/3}(v+2)h^{4/3}(v+4)\Psi_{4}(l_1,l_2,v;\phi)\nonumber\\
&&-\f{1}{2\lambda^2}V_cv^{8/3}h^{8/3}(v)\Psi_0(l_1,l_2,v;\phi)\nonumber\\
&&-\f{i\vartheta}{2\lambda}(16V_c)^{2/3}h^2(v)h^2(v-2)\Psi_{-2}(l_1,l_2,v;\phi)\nonumber\\
&&+\f{i\vartheta}{2\lambda}(16V_c)^{2/3}h^2(v)h^2(v+2)\Psi_{2}(l_1,l_2,v;\phi)\nonumber\\
&&+2^{13/3}\vartheta^2(1+\gamma^2)V_c^{1/3}A(v)h^4(v)\bigg(l_1^5l_2l_3+l_2^5l_1l_3+l_3^5l_1l_2\nonumber\\
&&-2l_1^8l_2^8h^6(v)-2l1^8l_3^8h^6(v)-2l_2^8l_3^8h^6(v)\bigg)\Psi(l_1,l_2,v;\phi)\Bigg]\nonumber\\
&&
\eea
where
\bea
\Psi_{\pm4}(l_1,l_2,v;\phi)&=&\Psi(\f{v\pm2}{v}l_1,\f{v\pm4}{v\pm2}l_2,v\pm4;\phi)+\Psi(\f{v\pm4}{v\pm2}l_1,\f{v\pm4}{v\pm2}l_2,v\pm4;\phi)
\nonumber\\
&&+\Psi(\f{v\pm2}{v}l_1,l_2,v\pm4;\phi)+\Psi(\f{v\pm4}{v\pm2}l_1,l_2,v\pm4;\phi)\nonumber\\
&&+\Psi(l_1,\f{v\pm2}{v}l_2,v\pm4;\phi)+\Psi(l_1,\f{v\pm4}{v\pm2}l_2,v\pm4;\phi),
\eea
\bea
\Psi_{\pm2}(l_1,l_2,v;\phi)&=&\textrm{sgn}(v\pm2)A(v\pm2)\bigg(l_1^4l_2^4\Psi(l_1,l_2,v\pm2;\phi)\nonumber\\
&&+l_2^4l_3^4\Psi(\f{v\pm2}{v}l_1,l_2,v\pm2;\phi)
+\Psi(l_1,\f{v\pm2}{v}l_2,v\pm2;\phi)\bigg)\nonumber\\
&&+A(v)\bigg(l_1^4l_2^4\Psi(l_1,l_2,v\pm2;\phi)+l_2^4l_3^4\Psi(\f{v\pm2}{v}l_1,l_2,v\pm2;\phi)
\nonumber\\
&&+\Psi(l_1,\f{v\pm2}{v}l_2,v\pm2;\phi)\bigg),
\eea
\bea
\Psi_{0}(l_1,l_2,v;\phi)&=&\chi_{v-2}A(v-2)(v-2)^{2/3}h^{2/3}(v-2)\bigg(\Psi(\f{v-2}{v}l_1,\f{v}{v-2}l_2,v;\phi)\nonumber\\
&&+\Psi(\f{v}{v-2}l_1,\f{v-2}{v}l_2,v;\phi)+\Psi(\f{v-2}{v}l_1,l_2,v;\phi)\nonumber\\
&&+\Psi(\f{v}{v-2}l_1,l_2,v;\phi)+\Psi(l_1,\f{v-2}{v}l_2,v;\phi)+\Psi(l_1,\f{v}{v-2}l_2,v;\phi)\bigg)\nonumber\\
&&+A(v+2)(v+2)^{2/3}h^{2/3}(v+2)\bigg(\Psi(\f{v+2}{v}l_1,\f{v}{v+2}l_2,v;\phi)\nonumber\\
&&+\Psi(\f{v}{v+2}l_1,\f{v+2}{v}l_2,v;\phi)+\Psi(\f{v+2}{v}l_1,l_2,v;\phi)\nonumber\\
&&+\Psi(\f{v}{v+2}l_1,l_2,v;\phi)+\Psi(l_1,\f{v+2}{v}l_2,v;\phi)+\Psi(l_1,\f{v}{v+2}l_2,v;\phi)\bigg),\nonumber\\
&&
\eea
and $\chi_a$ is a step function, it is zero when $a\leq0$ otherwise it is 1.

As an interesting 
result, because of the presence of negative powers of $p_i$ in the Hamiltonian constraint and the fact that the operator $\hat p_i^a$ is not the inverse of the operator $\widehat{p_i^{-a}}$ where $a$ is a positive real number, the quantum theory of the closed FLRW model (connection based quantization which was described in \cite{CK-2,CK-4}), is not a reduced theory of the quantum Bianchi IX that he have constructed. In retrospect, this result is not entirely unexpected, since we are employing a different strategy to deal with the inverse  corrections. One might still ask whether there is a quantization prescription where one can recover the $k$=1 FLRW model. In the next section, 
we consider the corresponding effective theories and their consequences.  As we shall see,
the effective theory suggests that there is such a prescription, but it is not the most natural choice.
We shall further discuss some of its properties.

\section{Effective Equations with Inverse Triad Corrections}
\label{sec:4}

By choosing the eigenvalues of the operators, negative powers of $|p_i|$ and $\hat A$ as corrections to the effective Hamiltonian, the modified effective Hamiltonian, with a generic matter density, is given by
\bea
\label{H-BIX}
\mathcal{H}_{\textrm BIX}&=&-\frac{V^4A(V)h^6(V)}{8\pi GV_c^6\gamma^2\lambda^{2}}\bigg(\sin\bar\mu_1c_1\sin\bar\mu_2c_2+\sin\bar\mu_1c_1\sin\bar\mu_3c_3
+\sin\bar\mu_2c_2\sin\bar\mu_3c_3\bigg)\nonumber\\
& &+\frac{\vartheta A(V)h^4(V)}{4\pi GV_c^4\gamma^2\lambda}\bigg(p_1^2p_2^2\sin\bar\mu_3c_3
+p_2^2p_3^2\sin\bar\mu_1c_1 +p_1^2p_3^2\sin\bar\mu_2c_2\bigg)\nonumber\\
& &-\frac{\vartheta^2(1+\gamma^2)A(V)h^4(V)}{8\pi GV_c^4\gamma^2}\bigg(2V[p_1^2+p_2^2+p_3^2]\nonumber\\
&&-\bigg[(p_1p_2)^{4}+(p_1p_3)^{4}+(p_2p_3)^{4}\bigg]\frac{h^6(V)}{V_c^6}\bigg)+\rho V \approx 0
\eea
where $A(V)$ and $h(V)$ are the same as Eqs.(\ref{Avl}),(\ref{hvl}) but now as a function of volume instead of the eigenvalue $v$.
\be
A(V)=\f{1}{V_c}(V+V_c-|V-V_c|)=
\left\{\begin{array}{lr}V/V_c & V<V_c\\1 & V\geq V_c\end{array}\right. ,
\ee
and
\be
h(V)=\sqrt{V+V_c}-\sqrt{|V-V_c|}.
\ee
Recall that $V_c=2\pi G\hbar\gamma\lambda$ sets the scale where the quantum effects are kicking in and change the qualitative behaviour of the equations.

The equation of motion for $p_1$ and $c_1$ are
\be
\dot p_1=\frac{1}{V_c^4\gamma}A(V)Vh^4(V)\cos\bar\mu_1c_1\bigg[\frac{V^2h^2(V)p_1}{V_c^2\lambda}(\sin\bar\mu_2c_2+\sin\bar\mu_3c_3)
-2\vartheta p_2p_3\bigg],
\ee
and
\be
\begin{split}
\dot c_1=&-\frac{h^5(V)}{V_c^6\gamma\lambda^{2}}\bigg(2p_2^2p_3^2p_1A(V)h(V)+V^4A_{,p_1}h(V)+6V^4A(V)h_{,p_1}\bigg)\\
&\times\bigg(\sin\bar\mu_1c_1\sin\bar\mu_2c_2+\sin\bar\mu_1c_1\sin\bar\mu_3c_3
+\sin\bar\mu_2c_2\sin\bar\mu_3c_3\bigg)\\
&+\frac{2\vartheta}{V_c^4\gamma\lambda}h^3(V)\bigg[\bigg(2p_1p_2^2A(V)h(V)+p_1^2p_2^2A_{,p_1}h(V)+4p_1^2p_2^2A(V)h_{,p_1}\bigg)\sin\bar\mu_3c_3\\
&+\bigg(2p_1p_3^2A(V)h(V)+p_1^2p_3^2A_{,p_1}h(V)+4p_1^2p_3^2A(V)h_{,p_1}\bigg)
\sin\bar\mu_2c_2\\
&+\bigg(p_2^2p_3^2A_{,p_1}h(V)+4p_2^2p_3^2A(V)h_{,p_1}\bigg)\sin\bar\mu_1c_1\bigg]\\
&-A(V)h^4(V)\frac{c_1\cos\bar\mu_1c_1}{2V_c^4\gamma}\left(\frac{V^3h^2(V)}{V_c^2\lambda}
(\sin\bar\mu_2c_2+\sin\bar\mu_3c_3)-2\frac{p_2^{3/2}p_3^{3/2}}{p_1^{1/2}}\vartheta\right)\\
&+A(V)h^4(V)\frac{c_2\cos\bar\mu_2c_2}{2V_c^4\gamma}\left(\frac{p_2^{5/2}p_3^{3/2}p_1^{1/2}h^2(V)}{V_c^2\lambda}
(\sin\bar\mu_1c_1+\sin\bar\mu_3c_3)-2\vartheta p_3V\right)\\
&+A(V)h^4(V)\frac{c_3\cos\bar\mu_3c_3}{2V_c^4\gamma}\left(\frac{p_3^{5/2}p_2^{3/2}p_1^{1/2}h^2(V)}{V_c^2\lambda}
(\sin\bar\mu_1c_1+\sin\bar\mu_2c_2)-2\vartheta p_2V\right)\\
&-\frac{\vartheta^2(1+\gamma^2)}{V_c^4\gamma}h^3(V)\Bigg[4p_1A(V)Vh(V)\\
&+(p_1^2+p_2^2+p_3^2)\left(\sqrt{\frac{p_2p_3}{p_1}}A(V)h(V)+8A(V)Vh_{,p_1}+2A_{,p_1}Vh(V)\right)\\
&-\frac{4}{V_c^6}p_1^{3}h^7(V)A(V)(p_2^{4}+p_3^{4})\\
&-\frac{1}{V_c^6}\bigg(10h^6(V)h_{,p_1}A(V)+h^7(V)A_{,p_1}\bigg)\bigg(p_1^{4}p_2^{4}+p_1^{4}p_3^{4}+p_2^{4}p_3^{4}\bigg)\Bigg]\\
&+4\pi G\gamma\bigg(V\frac{\partial\rho}{\partial p_1}+\rho\frac{1}{2}\sqrt{\frac{p_2p_3}{p_1}}\bigg)
\end{split}
\ee
where the partial derivatives of $A(V)$ and $h(V)$ respect to $p_i$ are
\begin{equation}
A_{,p_i}=\left\{\begin{array}{lr}\frac{1}{V_c}\sqrt{\frac{p_jp_k}{p_i}} & V<V_c\\ 0 & V>V_c \end{array} \right. \textrm{   ,   } i\neq j\neq k\neq i
\end{equation}
and
\be
\widehat h_{,p_i}=\frac{\sqrt{p_jp_k}}{4\sqrt{p_i}}\left[(V+V_c)^{-1/2}-\frac{\sqrt{|V-V_c|}}{V-V_c}\right]
\textrm{   ,   } i\neq j\neq k\neq i
\ee
The equations for $\dot p_2$,$\dot c_2$,$\dot p_3$ and $\dot c_3$ can be obtained by appropriate permutations.
One of the quantity which we want to know whether it is bounded or not is the expansion \cite{cs:unique},
\be
\begin{split}
\theta=&\frac{1}{2}\sum_i\frac{\dot p_i}{p_i}=\frac{1}{2V_c^4\gamma}A(V)Vh^4(V)\Bigg(\cos\bar\mu_1c_1\bigg[\frac{V^2h^2(V)}{V_c^2\lambda}(\sin\bar\mu_2c_2+\sin\bar\mu_3c_3)
-2\vartheta\frac{p_2p_3}{p_1}\bigg]\\
&+\cos\bar\mu_2c_2\bigg[\frac{V^2h^2(V)}{V_c^2\lambda}(\sin\bar\mu_1c_1+\sin\bar\mu_3c_3)
-2\vartheta\frac{p_1p_3}{p_2}\bigg]\\
&+\cos\bar\mu_3c_3\bigg[\frac{V^2h^2(V)}{V_c^2\lambda}(\sin\bar\mu_1c_1+\sin\bar\mu_2c_2)
-2\vartheta\frac{p_1p_2}{p_3}\bigg]\Bigg)
\label{tbix1}
\end{split}
\ee
At the limit of large volumes, the first term of $\dot p_1/p_1$ behaves like a combination of some trigonometric functions and the second as $\sqrt{p_2p_3}/p_1^{3/2}$. In the limit of small volume the first term behaves like $V^{10}$ and the second as $V^4p_2^2p_3^2$. For any value that the volume takes, $p_1$ or $p_2$ or $p_3$ can be small or arbitrary large numbers; therefore $\dot p_1/p_1$ is not bounded. Since there are similar statements for $\dot p_2/p_2$ and $\dot p_3/p_3$, the expansion is unbounded.

The other interesting geometrical observables to consider are shear and matter density. The shear is
given by
\be
\sigma^2=\frac{1}{3}[(H_1-H_2)^2+(H_2-H_3)^2+(H_1-H_3)^2]
\ee
where $H_i=1/2[\dot p_j/p_j+\dot p_k/p_k-\dot p_i/p_i] \textrm{ }(i\neq j\neq k)$ and $H_i-H_j=\dot p_j/p_j-\dot p_i/p_i$.
 and the expression for matter density in this modified theory is given by
\be
\label{dens-eq}
\begin{split}
\rho=&\frac{V^3A(V)h^6(V)}{8\pi GV_c^6\gamma^2\lambda^{2}}\left(\sin\bar\mu_1c_1\sin\bar\mu_2c_2+\sin\bar\mu_1c_1\sin\bar\mu_3c_3
+\sin\bar\mu_2c_2\sin\bar\mu_3c_3\right)\\
&-\frac{\vartheta A(V)}{4\pi GV_c^4\gamma^2\lambda}\left(\frac{p_1^{3/2}p_2^{3/2}}{p3^{1/2}}\sin\bar\mu_3c_3+\frac{p_2^{3/2}p_3^{3/2}}{p_1^{1/2}}\sin\bar\mu_1c_1+
\frac{p_1^{3/2}p_3^{3/2}}{p_2^{1/2}}\sin\bar\mu_2c_2\right)h^4(V)\\
&+\frac{\vartheta^2(1+\gamma^2)A(V)h^4(V)}{8\pi GV_c^4\gamma^2}\bigg(2\bigg[p_1^2+p_2^2+p_3^2\bigg]
-\bigg[\frac{(p_1p_2)^{7/2}}{p_3^{1/2}}+\frac{(p_1p_3)^{7/2}}{p2^{1/2}}\\
&+\frac{(p_2p_3)^{7/2}}{p_1^{1/2}}\bigg]\frac{h^5(V)}{V_c^5}\bigg)
\end{split}
\ee

With the same arguments which we used to prove unboundedness of the expansion, we can show that the shear and the density are unbounded, too.
Furthermore, it is easy to see that because of the function $A(V)$, when volume goes to zero, the
expansion, shear and density go to zero, too.

\begin{figure}
        \centering
        \begin{subfigure}
                \centering
                \includegraphics[width=0.5\textwidth, height=0.45\linewidth]{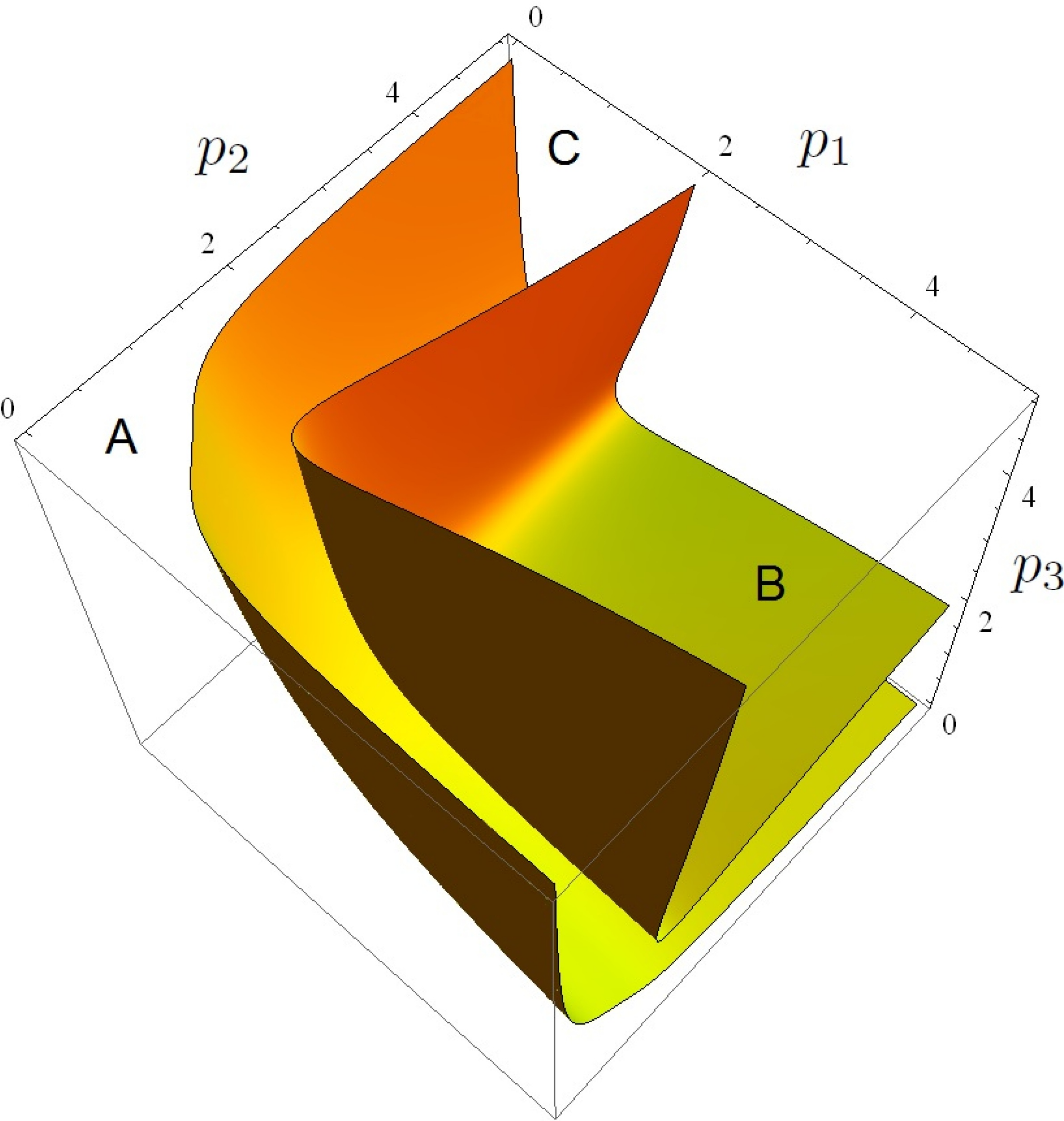}
        \end{subfigure}
     \begin{subfigure}
                \centering
                \includegraphics[width=0.45\textwidth, height=0.45\linewidth]{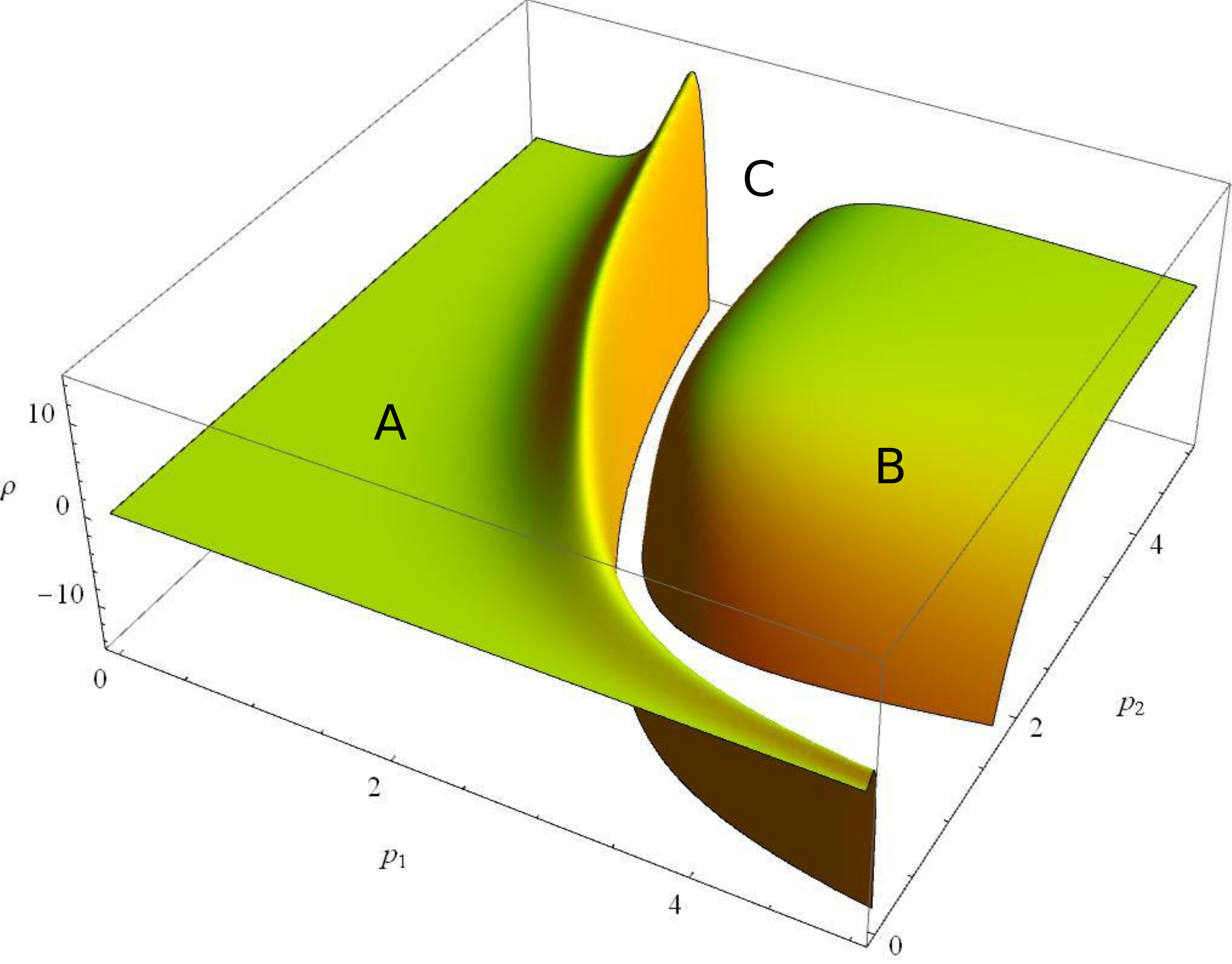}
        \end{subfigure}
\caption{Left, zero surfaces of
the maximum allowed density.  Right, maximum allowed density vs. $p_1$ and $p_2$ where $p_3=p_1$. Both in Planck units.}
\label{dens} \label{dens1}
\end{figure}

In the general case, as it can be seen in Fig.\ref{dens}, the maximum allowed density which arises from 
the modified Hamiltonian, has two distinct disconnected regions with positive values,
unlike the maximum allowed density in previous section which is always positive.
If we impose the weak energy condition and start the evolution within one region,
the universe cannot reach the other region. These two regions have different dynamics.
To study the vacuum Bianchi IX, we start from large volumes which lie in region B of Fig.\ref{dens} and, as we go to smaller volumes we cannot reach  zero volume because `crossing' to region A is not allowed. Therefore, there is a smallest reachable volume in region B and, since very large anisotropies are not allowed near this smallest volume, and the modified potential is not too large there, then we have, at most, finite oscillations before reaching the bounce. On the other hand, in the internal region A, the anisotropies are very large when some of the $p_i$ are very small, and then the volume of the universe cannot be large enough to start the evolution from there \cite{CKM}.

In \cite{gupt:singh} the authors used a modified effective equation and calculated the matter density with almost similar behaviour. The differences between their work and ours are:
i) The effect of operator $\hat A$ is neglected and ii) The matter density is defined as the eigenvalue of density operator which was defined as $\hat\rho:=-\widehat{V^{-1}}\hat{\mathcal H_{grav}}$, while we use the more standard definition of matter density which is $\rho:=-\mathcal H_{grav}/V$. Furthermore, they showed that if one calculates the matter density in the usual manner, the maximum allowed density in their modified theory behaves more similar to the density in original Bianchi IX effective theory of \cite{Ed-BIX},  than the matter density in Eq.(\ref{dens-eq}).

Although the effective Hamiltonian without inverse triad correction \cite{Ed-BIX} reduces to the effective Hamiltonian for  closed FLRW model with less correction \cite{closed}, because of the presence of the effects from the operators $\widehat{|p_i|^{-1/4}}$ and its positive powers in the curvature term of the Hamiltonian, this property no longer holds for the modified Hamiltonian Eq.(\ref{H-BIX}) and the closed FLRW model effective Hamiltonian with inverse triad correction is not a reduction of this modified Hamiltonian. 


\subsection{A different choice: Bianchi IX reduces to  FLRW k=1}

As we mentioned before, the Bianchi IX effective theory with inverse triad corrections does not reduce to the closed FLRW model.
However, by keeping the effects of operator $\hat A$ and neglecting the other corrections in the gravitational part of the Hamiltonian constraint, one can construct a Bianchi IX effective theory which has some part of the inverse corrections and it {\it does} reduce to the closed FLRW model with inverse triad corrections.
In this model, the Hamiltonian constraint is given by
\be
\begin{split}
\mathcal{H}_{eff}=&-\frac{V}{8\pi G\gamma^2\lambda^2}A(V)(\sin\bar\mu_1c_1\sin\bar\mu_2c_2+\sin\bar\mu_1c_1\sin\bar\mu_3c_3
+\sin\bar\mu_2c_2\sin\bar\mu_3c_3)\\
&+\frac{\sigma}{8\pi G\gamma^2\lambda}A(V)\left(\frac{p_1p_2}{p_3}\sin\bar\mu_3c_3+
\frac{p_2p_3}{p_1}\sin\bar\mu_1c_1+\frac{p_1p_3}{p_2}\sin\bar\mu_2c_2\right)\\
&-\frac{\sigma^2(1+\gamma^2)}{32\pi G\gamma^2}A(V)\left(2\frac{p_1^{3/2}}{\sqrt{p_2p_3}}+2\frac{p_2^{3/2}}{\sqrt{p_1p_3}}+2\frac{p_3^{3/2}}{\sqrt{p_1p_2}}
-\frac{(p_1p_2)^{3/2}}{p_3^{5/2}}-\frac{(p_1p_3)^{3/2}}{p_2^{5/2}}-\frac{(p_2p_3)^{3/2}}{p_1^{5/2}}\right)\\
&+U(\phi)V+\frac{1}{2V_c^6}p_\phi^2V^2h(V)^6
\end{split}
\ee
\be
\dot p_1=\frac{1}{\gamma\lambda}A(V)\cos\bar\mu_1c_1\left[p_1(\sin\bar\mu_2c_2+\sin\bar\mu_3c_3)
-\lambda\sigma\sqrt{\frac{p_2p_3}{p_1}}\right]
\ee
\be
\begin{split}
\dot{c_1}=&-\frac{1}{\gamma\lambda^2}\left(\frac{1}{2}\sqrt{\frac{p_2p_3}{p_1}}A(V)+A_{,p_1}V\right)\left(\sin\bar\mu_1c_1\sin\bar\mu_2c_2+\sin\bar\mu_1c_1\sin\bar\mu_3c_3+\sin\bar\mu_2c_2\sin\bar\mu_3c_3\right)\\
&-\frac{\sigma}{\lambda\gamma}\Bigg(\bigg[\frac{p_2p_3}{p_1^2}A(V)-\frac{p_2p_3}{p_1}A_{,p_1}\bigg]\sin\bar\mu_1c_1-\bigg[\frac{p_3}{p_2}A(V)+\frac{p_1p_3}{p_2}A_{,p_1}\bigg]\sin\bar\mu_2c_2
\\
&-\bigg[\frac{p_2}{p_3}A(V)+\frac{p_1p_2}{p_3}A_{,p_1}\bigg]\sin\bar\mu_3c_3\Bigg)\\
&-A(V)\frac{c_1\cos\bar\mu_1c_1}{2\gamma\lambda}\left(-\frac{\lambda\sigma}{p_1}\sqrt{\frac{p_2p_3}{p_1}}+\sin\bar\mu_2c_2+\sin\bar\mu_3c_3\right)\\
&+A(V)\frac{c_2\cos\bar\mu_2c_2}{2\gamma\lambda\sqrt{p_1}}\left(-\lambda\sigma\sqrt{\frac{p_3}{p_2}}+\frac{p_2}{\sqrt{p_1}}\sin\bar\mu_1c_1+\frac{p_2}{\sqrt{p_1}}\sin\bar\mu_3c_3\right)\\
&+A(V)\frac{c_3\cos\bar\mu_3c_3}{2\gamma\lambda\sqrt{p_1}}\left(-\lambda\sigma\sqrt{\frac{p_2}{p_3}}+\frac{p_3}{\sqrt{p_1}}\sin\bar\mu_1c_1+\frac{p_3}{\sqrt{p_1}}\sin\bar\mu_2c_2\right)\\
&-A(V)\frac{\sigma^2(1+\gamma^2)}{4\gamma}\left(\frac{5}{2}\frac{p_2^{3/2}p_3^{3/2}}{p_1^{7/2}}-\frac{1}{p_1^{3/2}\sqrt{p_2p_3}}(p_2^2+p_3^2)-\frac{3\sqrt{p_1}}{2}(\frac{p_2^{3/2}}{p_3^{5/2}}+\frac{p_3^{3/2}}{p_2^{5/2}})+3\sqrt{\frac{p_1}{p_2p_3}}\right)\\
&+8\pi G\gamma\sqrt{\frac{p_2p_3}{p_1}}U(\phi)+\frac{4\pi G\gamma}{V_c^6}p_\phi^2h^5(V)\bigg[p_2p_3h(V)+6V^2h_{,p_1}\bigg]
\end{split}
\ee
The expansion is
\be
\begin{split}
\theta=&\frac{1}{2\gamma\lambda}A(V)(\cos\bar\mu_1c_1\left[(\sin\bar\mu_2c_2+\sin\bar\mu_3c_3)
-\lambda\sigma\frac{1}{p_1}\sqrt{\frac{p_2p_3}{p_1}}\right]\\
&+\cos\bar\mu_2c_2\left[(\sin\bar\mu_1c_1+\sin\bar\mu_3c_3)
-\lambda\sigma\frac{1}{p_2}\sqrt{\frac{p_1p_3}{p_2}}\right]
+\cos\bar\mu_3c_3\left[(\sin\bar\mu_1c_1+\sin\bar\mu_2c_2)
-\lambda\sigma\frac{1}{p_3}\sqrt{\frac{p_1p_2}{p_3}}\right])
\end{split}
\ee
For large volume, the second term of $\dot p_1/p_1$ behaves like $\sqrt{p_2p_3}/p_1^{3/2}$ and in the limit of small volume behaves like $p_2p_3/p_1$. Since $p_i$'s can be small or large numbers, there is no bound for $\dot p_1/p_1$ and with similar arguments about unboundedness of $\dot p_2/p_2$ and $\dot p_3/p_3$, it can be proved that expansion is not bounded.\\

\section{discussion}
\label{sec:5}

This paper is the first in a series devoted to the study of the Bianchi IX model within LQC. In this contribution we
introduced for the first time inverse corrections for Bianchi IX and explored some of its properties. In particular we have studied the effective theory that follows from the quantum theory and have considered the behaviour of several geometric scalars. This is important to
study singularity resolution within LQC. Some of these questions are explored in the second paper of this series \cite{CM-BIX}, where numerical solutions to the effective equations with a massless scalar field are studied. 

In the study of the behaviour of expansion and shear for the effective theory, we have shown that  these scalars are not absolutely bounded, which might be a signal that the quantization is problematic \cite{cs:unique}. However, when one takes into account some energy conditions, one learns that the allowed region where solutions to the effective equations can be, becomes disconnected. There are two allowed regions and the solutions have to lie within one of them. The region where the would be singularity lies, and where large anisotropies are allowed, is disconnected from the region with large volume. Thus, any realistic universe that reaches large volume at recollapse can not reach that region, and it can, at most, have a finite number of oscillations. Thus, one might expect that loop quantum corrections to the dynamics have an  important effect on  the avoidance, not only of the singularity, but of the mixmaster behaviour that is so characteristic of the classical dynamics. This and other issues will be studied in more detail in the third paper of the series \cite{CKM-2}.

\section*{Acknowledgements}

This work was in part supported by DGAPA-UNAM IN103610 grant, by CONACyT 0177840 
and 0232902 grants, by the PASPA-DGAPA program, by NSF
PHY-1505411 and PHY-1403943 grants, and by the Eberly Research Funds of Penn State.

\end{document}